\documentclass[12pt]{article}
\newenvironment{wileykeywords}{\textsf{Keywords:}\hspace{\stretch{1}}}
{\hspace{\stretch{1}}\rule{1ex}{1ex}}
\usepackage{amsmath,amssymb}
\usepackage[dvipdfmx]{graphicx}
\usepackage{caption}
\usepackage{color}
\usepackage{dcolumn}
\usepackage{bm}
\usepackage[numbers,super,comma,sort&compress]{natbib}
\usepackage[nolists, nomarkers, figuresfirst]{endfloat}
\definecolor{background-color}{gray}{0.98}
\usepackage{float}
\newcommand{\circled}[1]{\raise0.2ex\hbox
{\textcircled{\scriptsize{#1}}}}
\usepackage[none]{hyphenat}
\title{Path Integral Brownian Chain Molecular Dynamics: A Simple
 Approximation of Quantum Vibrational Dynamics}

\author{
  Motoyuki Shiga\thanks{
  Center for Computational Science and e-Systems,
  Japan Atomic Energy Agency, 178-4-4, Wakashiba, Kashiwa, Chiba,
  277-0871, Japan}
}

\begin{document}

\maketitle

\begin{abstract}

An approximate approach to quantum vibrational dynamics,
 ``Brownian Chain Molecular Dynamics (BCMD)'', is proposed
 to alleviate the chain resonance and curvature problems
 in the imaginary time-based path integral (PI) simulation.
Here the non-centroid velocity is randomized at each step when solving
 the equation of motion of path integral molecular dynamics.
This leads to a combination of the Newton equation and the overdamped
 Langevin equation for the centroid and non-centroid variables,
 respectively.
BCMD shares the basic properties of other PI approaches such
 as centroid and ring polymer molecular dynamics:
It gives the correct Kubo-transformed correlation function at
 short times, conserves the time symmetry, has the correct
 high-temperature/classical limits, gives exactly
 the position and velocity autocorrelations of harmonic
 oscillator systems, and does not have the zero-point leakage problem.
Numerical tests were done on
 simple molecular models and liquid water.
On-the-fly ab initio BCMD simulations were performed for
 the protonated water cluster,
 H$_5^{}$O$_2^+$, and its isotopologue, D$_5^{}$O$_2^+$.

\end{abstract}


\begin{wileykeywords}
 path integral simulations,
 molecular dynamics,
 vibrational dynamics,
 semiclassical theory,
 ab initio simulations
\end{wileykeywords}

\clearpage

\bibliographystyle{jpc}
\renewcommand{\baselinestretch}{1.5}
\normalsize

\begin{figure}[h]
\centering
\colorbox{background-color}{
\fbox{
\begin{minipage}{1.0\textwidth}
\includegraphics[width=50mm,height=50mm]{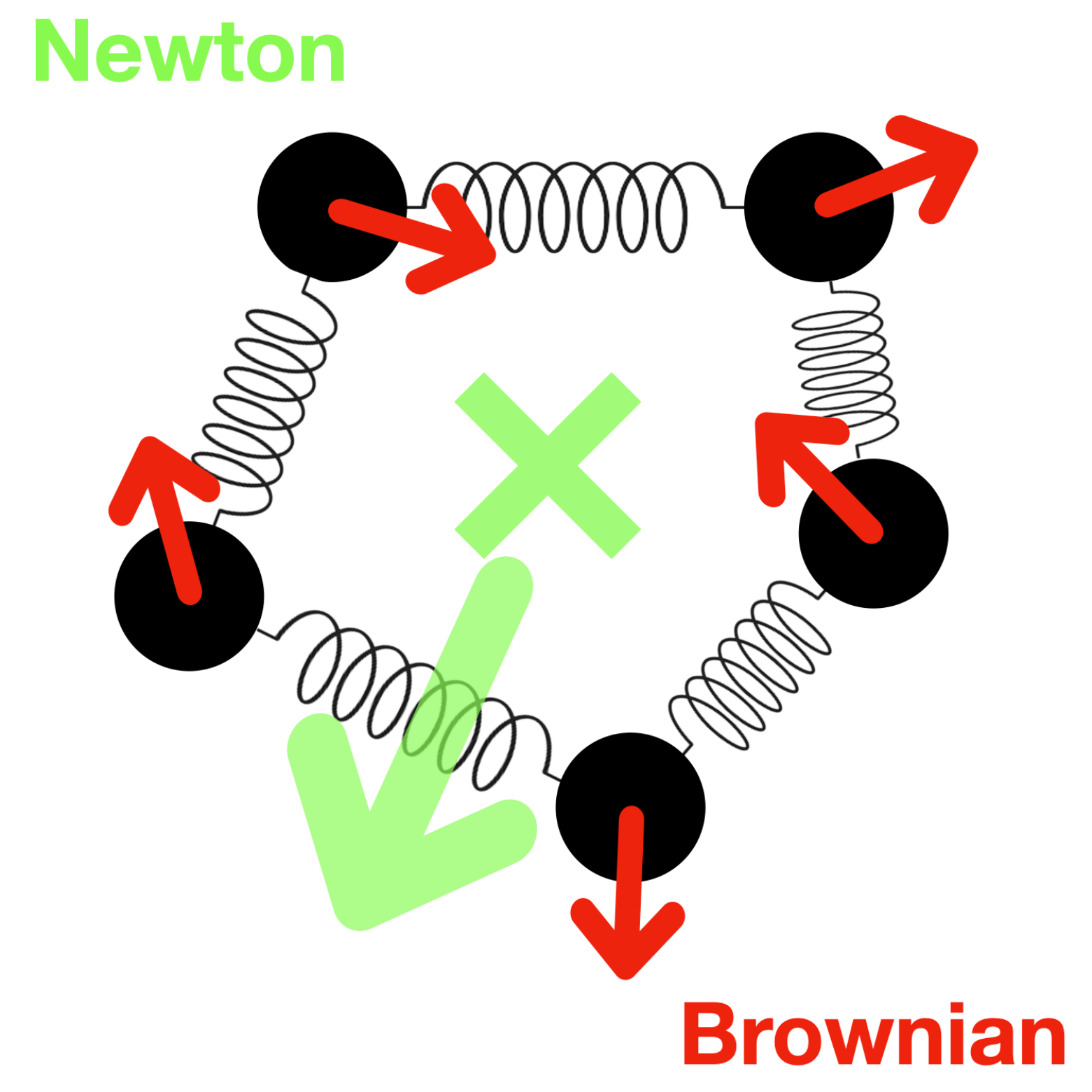}
\\
An approximate approach to quantum vibrational dynamics,
 ``Brownian Chain Molecular Dynamics'', is proposed to alleviate
 the chain resonance and curvature problems in the imaginary
 time-based path integral simulation. Here the non-centroid
 velocity is randomized at each step when solving the equation
 of motion of path integral molecular dynamics. This leads to
 a combination of the Newton equation and the overdamped
 Langevin equation for the centroid and non-centroid variables,
 respectively.
\end{minipage}
}}
\end{figure}

\makeatletter
\renewcommand\@biblabel[1]{#1.}
\makeatother

\clearpage

\section{\sffamily \large INTRODUCTION \label{sec1}}

Various semiclassical methods have been proposed as approximate
 solutions to quantum vibrational dynamics of many-body systems.
 \cite{cao1994formulation,miller2001semiclassical,shi2003relationship,nakayama2003forward,craig2004quantum,poulsen2005static,hyeon2009semiquantum,hele2015boltzmann,liu2016path,rossi2018fine,trenins2019path,kapil2020inexpensive}
Centroid molecular dynamics (CMD)
 \cite{cao1994formulation,jang1999derivation}
 and the ring polymer molecular dynamics (RPMD)
 \cite{craig2004quantum,braams2006short}
 are among such representative methods in the
 framework of path integral (PI) simulations.
They could be categorized as an extension of path integral molecular
 dynamics (PIMD) simulations that utilize the isomorphism between
 the quantum fluctuation of a particle and that of a cyclic chain
 of its classical replica (beads).
The family of PI simulations is based on quantum statistical
 mechanics and is rigorous for time-independent physical
 properties in the limit of infinite number of beads.
In the CMD and RPMD methods, the time correlation function
 from the classical dynamics of cyclic chains is regarded as
 an approximation of the Kubo-transformed quantum
 correlation function.
These CMD and RPMD correlation functions are correct
 at short times, preserves time symmetry, follows basic
 conservation laws, and has the correct
 high temperature/classical limits.
In addition, the position and velocity autocorrelation functions
 are exact for harmonic oscillators.
Since the thermodynamic equilibrium is maintained
%
in the
%
 CMD and RPMD trajectories, so-called
 the ``zero-point leakage'' problem
 in semiclassical methods is not present in CMD and RPMD\@.
Accordingly, the CMD and RPMD methods have been applied
 successfully in a variety of systems.


However, the semiclassical methods always have some limitations.
A common drawback of CMD and RPMD is in the vibrational spectra
 of high frequency modes.
As the temperature is lower, the CMD curvature problem
 \cite{witt2009applicability,ivanov2010communications}
 and the RPMD chain resonance problem
 \cite{shiga2008ab,witt2009applicability} become more prominent.
The former arises from a spurious coupling between rotations
 and vibrations due to the adiabatic separation of the
 centroid variable and the non-centroid variables.
The latter is due to spurious resonance between the vibrations
 of the physical mode and the vibrations of the cyclic chain.
These problems appear, for example, as peak shifts and peak splits
 in the OH and CH vibrations of water and methane, respectively,
 which do not go away even at room temperature.
The thermostatted RPMD (TRPMD)
 method was later proposed to suppress chain resonances,
 where the RPMD equation of motion
 is coupled to the Langevin thermostats. \cite{rossi2014remove}
The path integral Liouville dynamics (PILD) was also proposed to
 solve these problems at a cost of computing the Hessian matrix.
 \cite{liu2016path}


In this study, we propose a simple method that completely
 eliminates the chain resonance problem of RPMD\@.
This method does not eliminate the curvature problem,
 but it does reduce the severity of the problem with
 respect to spurious frequency shifts over CMD\@.
It uses the same type of equation of motion used
 in the family of path integral simulation
 techniques (PIMD, CMD, RPMD).
The main difference from other methods is
 that the non-centroid velocities are randomized
 at each step according to the Maxwell-Boltzmann
 distribution.
It turns out that
 this is equivalent to solving the Newtonian dynamics
 for centroid variables and the Brownian
 (overdamped Langevin) dynamics for
 non-centroid variables at the same time.
For this reason, this method is named here as
 ``Brownian Chain Molecular Dynamics (BCMD)''.
While randomization of the non-centroid velocities can
 eliminate the chain resonance problem,
 the nonadiabaticity between the centroid
 and non-centroid variables can
 alleviate the curvature problem.
Yet BCMD shares the basic nature of CMD and RPMD;
the Kubo-transformed correlation function is
 correct at short times, has the correct time symmetry,
 has the correct high-temperature/classical limits,
 provides exact results for the position
 and velocity autocorrelations for harmonic
 oscillators, and does not have the zero-point leakage problem.


In Section \ref{sec2}, the theory and numerical algorithm
 of the BCMD method are described.
In Section \ref{sec3}, the short-time behavior of BCMD
 is tested for one-dimensional oscillator model.
The vibrational spectra of BCMD are tested for gaseous
 OH molecule,
 and 
 H$_2^{}$O molecule
 and its isotopologues
 and liquid water isotopologues.
BCMD results are compared with the available data for MD, CMD,
 RPMD and TRPMD\@.
The ab initio BCMD simulations are demonstrated for
 H$_5^{}$O$_2^+$ and D$_5^{}$O$_2^+$ with
 the on-the-fly electronic structure calculations
 based on the second-order M{\o}ller-Plesset perturbation
 theory (MP2).
The infrared spectra obtained from the dipole
 autocorrelation function are compared with
 experimental results.
In Section \ref{sec4}, the conclusive remarks are given.

\section{\sffamily \large THEORY \label{sec2}}

\subsection{\sffamily \large Equation of motion in PI methods \label{sec2.1}}

The imaginary time path integral theory is a formulation for the
 statistical mechanics of quantum many-body systems.
 \cite{feynman2010quantum,feynman1972statistical,schulman2012techniques}
Let us consider the simple case of a single quantum particle
 moving in a one-dimensional potential energy
 surface, $V(x)$, with mass $m$.
Extension to multidimensional systems is
 described in Section \ref{sec2.6}\@.
The quantum partition function of the canonical ensemble
 at temperature $T$ can be expressed as
\begin{eqnarray}
 Z = \lim_{P\rightarrow\infty}
 \left(\frac{mP}{2\pi\beta\hbar^2}\right)^{\frac{P}{2}}
 \int d{\bf x} \exp\left(-\frac{S[{\bf x}]}{\hbar} \right)
 \label{eq1-1}
\end{eqnarray}
 where $\beta \equiv \frac{1}{k_{\rm B}^{}T}$ and
 $\hbar \equiv \frac{h}{2\pi}$ with the Boltzmann
 constant $k_{\rm B}^{}$ and the Planck constant $h$.
The functional
 $S[{\bf x}] \equiv \beta \hbar V_{\rm eff}^{}({\bf x})$
 represents the action with respect to a cyclic path
 ${\bf x} = (x_1^{}, \cdots, x_P^{})^T$
 along the imaginary time
 $\tau_j^{}=\frac{\beta\hbar{(j-1)}}{P}$,
 and
\begin{eqnarray}
 V_{\rm eff}^{}({\bf x})
 = \frac{1}{2} m \omega_P^2 {\bf x}^T{\bf A}{\bf x}
 + \phi({\bf x}).
 \label{eq1-2}
\end{eqnarray}
Equation (\ref{eq1-1}) is isomorphic to the classical partition function
 of a system of $P$ replicated particles (beads)
 with the effective potential
 of $V_{\rm eff}^{}({\bf x})$.
 \cite{chandler1981exploiting,parrinello1984study,tuckerman2010statistical}
In the rhs of Equation (\ref{eq1-2}), the first term is the
 harmonic interaction between
 the adjacent beads with the characteristic frequency
 $\omega_P^{} \equiv \frac{\sqrt{P}}{\beta\hbar}$.
The elements of the matrix ${\bf A}$ is defined as
 $A_{ij}^{} = 2\delta_{i,j}^{}-\delta_{i,j+1}^{}-\delta_{i,j-1}^{}$
 according to a cyclic boundary condition,
 $0\rightarrow P$ and $P+1\rightarrow 1$.
The second term is the bead average of the physical potential,
\begin{eqnarray}
 \phi({\bf x}) \equiv \frac{1}{P} \sum_{j=1}^{P} V(x_j^{}).
 \label{eq1-3}
\end{eqnarray}


The equations of motion for PI methods (PIMD, CMD, and RPMD)
 can be commonly derived from the Hamiltonian
 in normal mode space where
 the first term of Equation (\ref{eq1-2}) is diagonalized.
 \cite{witt2009applicability}
This is used in the BCMD method as well.
The normal mode transformation is obtained from
 the diagonalization of the ${\bf A}$ matrix,
 $P{\bf A}{\bf U} = \boldsymbol{\lambda}{\bf U}$,
 which can be solved analytically.
The eigenvalues are
\begin{eqnarray}
 &&
 \lambda_{2k}^{}
 = 4P\sin^2\left(\frac{\pi k}{P}\right)
 \ \ \ \ \mbox{($k\neq 0$)},
 \nonumber \\ &&
 \lambda_{2k+1}^{}
 = 4P\sin^2\left(\frac{\pi k}{P}\right)
 \ \ \mbox{($k \le \frac{P-1}{2}$)},
 \label{eq1-4}
\end{eqnarray}
and the eigenvectors are
\begin{eqnarray}
 &&
 U_{j,2k}^{} = 
 \sqrt{\frac{2}{P}} \cos\left(\frac{2 \pi j k}{P}\right)
 \ \ \ \ \mbox{($k\neq 0$)},
 \nonumber \\ &&
 U_{j,2k+1}^{} = 
 \begin{cases}
 \sqrt{\frac{1}{P}} & \mbox{($k=0$)} \\
 \sqrt{\frac{2}{P}} \sin\left(\frac{2 \pi j k}{P}\right)
 & \mbox{($k \le \frac{P-1}{2}$)}.
 \end{cases}
 \label{eq1-5}
\end{eqnarray}
Using the normal modes
\begin{eqnarray}
 {\bf q} = \frac{1}{\sqrt{P}} {\bf U}^T{\bf x}
 \label{eq1-6}
\end{eqnarray}
and its momentum conjugate ${\bf p}$,
the Hamiltonian is constructed as
\begin{eqnarray}
 H({\bf p},{\bf q})
 = \frac{1}{2}{\bf p}^T \boldsymbol{\mu}^{-1} {\bf p}
 + \frac{1}{2}m\omega_P^2 {\bf q}^T \boldsymbol{\lambda}{\bf q}
 + \phi({\bf x}({\bf q})),
 \label{eq1-7}
\end{eqnarray}
where $\boldsymbol{\mu}$ is a diagonal matrix, each of whose
 element corresponds to the fictitious mass of the normal mode,
 ${\rm diag}(\boldsymbol{\mu}) = (m,\mu_2^{},\cdots,\mu_P^{})$.
The canonical equation of motion derived from
 Equation (\ref{eq1-7}) is obtained as
\begin{eqnarray}
 \dot{\bf q} = \boldsymbol{\mu}^{-1} {\bf p}, \ \ \ \
 \dot{\bf p} = - m\omega_P^2 \boldsymbol{\lambda} {\bf q}
 - \nabla_{\bf q}^{} \phi({\bf x}({\bf q})).
 \label{eq1-8}
\end{eqnarray}
The first mode is the centroid variable,
 $q_1^{} = \frac{1}{P} \sum_{j=1}^P x_j^{}$,
 and $\lambda_1^{}=0$.
Thus, Equation (\ref{eq1-8}) is a set of equations
 for the centroid variable,
\begin{eqnarray}
 \dot{q}_1^{} = \frac{p_1^{}}{m} = v_1^{}, \ \ \ \
 \dot{p}_1^{} = - \frac{\partial \phi}{\partial q_1^{}},
 \label{eq1-9}
\end{eqnarray}
and the non-centroid variables,
\begin{eqnarray}
 \dot{q}_\alpha^{} = \frac{p_\alpha^{}}{\mu_\alpha^{}} = v_\alpha^{}, \ \ \ \
 \dot{p}_\alpha^{} = f_\alpha^{},
 \label{eq1-10}
\end{eqnarray}
where the non-centroid forces are
\begin{eqnarray}
 f_\alpha^{} = - m\omega_P^2 {\lambda}_\alpha^{} {q}_\alpha^{}
 - \frac{\partial \phi}{\partial q_\alpha^{}}.
 \label{eq1-11}
\end{eqnarray}
In Eqs.(\ref{eq1-9}) and (\ref{eq1-11}),
 the normal mode forces with respect to $\phi$
 can be obtained using the relation,
\begin{eqnarray}
 \frac{\partial \phi}{\partial q_\beta^{}}
 = \sqrt{P} \sum_{j=1}^P U_{j,\beta}^{}
 \frac{\partial \phi}{\partial x_j^{}}
 \ \ \ \ (1 \le \beta \le P).
 \label{eq1-12}
\end{eqnarray}
The non-centroid fictitious masses are set to
 $\mu_\alpha^{} = \lambda_\alpha^{} m$ in PIMD,
 $\mu_\alpha^{} = \gamma_{\rm cmd}^{-2} \lambda_\alpha^{} m$ in CMD
 (where $\gamma_{\rm cmd}^{} > 0$ is the adiabaticity parameter,
 which should be sufficiently large for a convergence),
 and $\mu_\alpha^{} = m$ for all $\alpha \ge 2$ in RPMD\@.
Temperature control also depends on the PI method.
Thermostats are used to control the temperature of
 both centroid and non-centroid variables in PIMD,
 only to the non-centroid variables in CMD and TRPMD,
 but thermostats are not used in RPMD\@.
%

\subsection{\sffamily \large BCMD \label{sec2.2}}

The BCMD method is based on the set of equations of motion,
 Eq.(\ref{eq1-9}) and a similar one to Eq.(\ref{eq1-10}) but with
 randomization of the non-centroid velocity,
 $v_\alpha^{}(t)=v_\alpha^{\prime}(t)$ for $\alpha \ge 2$.
In other words, at each step interval of $\Delta t$,
 $v_\alpha^{\prime}(t)$ is randomly sampled according
 to the Maxwell distribution with a variance of
\begin{eqnarray}
 \left\langle \left[ v_\alpha^{\prime}(t) \right]^2 \right\rangle
 = \frac{1}{\beta \mu_\alpha^{} }.
 \label{eq2-1}
\end{eqnarray}
Now let us show that this leads to
 Brownian dynamics, which is a generator of
 the canonical ensemble.
Applying the velocity-Verlet algorithm to Eq.(\ref{eq1-10}) with
 the initial non-centroid velocities $v_\alpha^{\prime}(t)$,
 the short-time evolution of the non-centroid modes
 from $t$ to $t + \Delta t$ is given by
\begin{eqnarray}
 q_\alpha^{}(t+\Delta t) = q_\alpha^{}(t)
 + v_\alpha^{\prime}(t)\Delta t
 + \frac{{f}_\alpha^{}(t)}{\mu_\alpha^{}}
 \frac{(\Delta t)^2}{2}.
 \label{eq2-2}
\end{eqnarray}
Introducing the notations
\begin{eqnarray}
 \gamma \equiv \frac{2}{\Delta t}
 \label{eq2-3}
\end{eqnarray}
and
\begin{eqnarray}
 {\xi}_\alpha^{}(t) \equiv \mu_\alpha^{}
 \gamma v_\alpha^{\prime}(t),
 \label{eq2-4}
\end{eqnarray}
Eq.(\ref{eq2-2}) can be rewritten in the form of
 the Brownian dynamics (overdamped Langevin equation),
\begin{eqnarray}
 \mu_\alpha^{} \gamma \dot{q}_\alpha^{}(t)
 = {f}_\alpha^{}(t) + {\xi}_\alpha^{}(t),
 \label{eq2-5}
\end{eqnarray}
where
\begin{eqnarray}
 \dot{q}_\alpha^{}
 = \frac{q_\alpha^{}(t+\Delta t) - q_\alpha^{}(t)}{\Delta t}.
 \label{eq2-6}
\end{eqnarray}
Combining to Eqs.(\ref{eq2-1}) and (\ref{eq2-4}), one obtains
\begin{eqnarray}
 \left\langle \left[{\xi}_\alpha^{}(t) \right]^2 \right\rangle
 = \frac{2\mu_\alpha^{} \gamma}{\beta}\frac{1}{\Delta t},
 \label{eq2-7}
\end{eqnarray}
indicating that ${\xi}_\alpha^{}(t)$ corresponds to
 a random force subject to the fluctuation-dissipation theorem,
\begin{eqnarray}
 {\xi}_\alpha^{}(t)
 = \sqrt{\frac{2\mu_\alpha^{} \gamma}{\beta}}{\zeta}(t),
 \label{eq2-8}
\end{eqnarray}
 with the white noise
 $\langle {\zeta}^{}(t) {\zeta}^{}(t^\prime) \rangle
 = \delta(t-t^\prime)$.
Therefore, the BCMD is the combination between the Newtonian dynamics
 with respect to the centroid variables, Eq.(\ref{eq1-9}),
 and the Brownian dynamics with respect to the non-centroid variables,
 Eq.(\ref{eq2-5}).


The major difference between BCMD and (T)RPMD is that the inertia
 term of (T)RPMD, i.e.,
 $m\ddot{q}_\alpha^{}$ where $m$ is the atomic mass,
 is not present in the non-centroid equation of motion, Eq.(\ref{eq2-5}).
The presence of the inertia term together with the spring
 force in the first term of the right-hand side of Eq.(\ref{eq2-5})
 is the source of the chain resonance problem.
In TRPMD the resonance can be diminished by applying a random
 force with the damping parameter being carefully optimized.
On the other hand, BCMD has no inertia term, so it can be used
 safely avoiding any resonant behavior from the outset,
 and this is of great advantage in the calculation of
 vibrational spectra using BCMD\@.
Comparing BCMD with RPMD and CMD, they are derived
 from the same equation of motion, Eq.(\ref{eq1-10}), and there
 is a clear correspondence among them in theory and
 numerical algorithm.
As discussed in Section  \ref{sec2.5}, BCMD has also a
 close relationship with path integral hybrid Monte Carlo
 (PIHMC) in terms of the numerical algorithm.


Eq.(\ref{eq1-10}) has turned into Eq.(\ref{eq2-5})
 simply by randomizing the non-centroid velocities
 with the Maxwell-Boltzmann distribution at each step
 in the velocity-Verlet algorithm;
Randomizing $v_\alpha^{}$ every step has changed
 a deterministic second-order differential equation
 of $q_\alpha^{}$ ({\it i.e.}, $\mu_\alpha^{}\ddot{q}_\alpha^{}=f_\alpha^{}$)
 into a stochastic first-order differential equation of $q_\alpha^{}$
 with respect to $t$ ({\it i.e.}, Eq.(\ref{eq2-5})).
As described in Eqs.(\ref{eq2-4}) and (\ref{eq2-8}),
 the random generation of velocities
 is the origin of the random force, $\xi_\alpha^{}$,
 with a white noise that obeys
 the fluctuation-dissipation theorem.
Thus it is guaranteed in the BCMD method thermal equilibrium
 of the system is maintained.
It is stressed once again that
 it is the nature of stochastic first-order differential equation
 of the BCMD method that ensures the system being completely free
 of the chain resonance behavior,
 in contrast to the case of the (T)RPMD method.

Now the mass of the non-centroid variables, $\mu_\alpha^{}$,
 has to be determined.
The smaller the $\mu_\alpha^{}$ value,
 the shorter the time scale of non-centroid variables
 would become, and the larger the adiabatic separation
 would be from the centroid variable.
The limiting case of $\mu_\alpha^{} \rightarrow 0$
 would simply correspond to CMD-like adiabatic dynamics.
Instead, let us herein determine the $\mu_\alpha^{}$ value
 from the short-time behavior of free particle system.
For the case of $V(x)=0$, Eq.(\ref{eq2-5}) becomes
\begin{eqnarray}
 \mu_\alpha^{} \gamma \dot{q}_\alpha^{} = 
 - \kappa_\alpha^{} {q}_\alpha^{} + {\xi}_\alpha^{},
 \label{eq2-9}
\end{eqnarray}
where
\begin{eqnarray}
 \kappa_\alpha^{} \equiv m \omega_P^2 {\lambda}_\alpha^{}.
 \label{eq2-10}
\end{eqnarray}
Equation (\ref{eq2-9}) corresponds exactly to the
 Ornstein-Uhlenbeck process. \cite{gardiner1985handbook}
Thus, the mean square displacement (MSD) is 
\begin{eqnarray}
 \left\langle \left| q_\alpha^{}(t)- q_\alpha^{}(0) \right|^2
 \right\rangle = \frac{2}{\kappa_\alpha^{}\beta}
 \left[1- \exp\left(-\frac{\kappa_\alpha^{}|t|}{\mu_\alpha^{}\gamma}\right) \right],
 \label{eq2-11}
\end{eqnarray}
 indicating that the non-centroid position will be
 uncorrelated after the relaxation time,
 $\tau_\alpha^{}=\frac{\mu_\alpha^{} \gamma}{\kappa_\alpha^{}}$.
Since it is expected that the imaginary time path integrals
 are not able to describe the quantum interference effects
 beyond the thermal time $\beta\hbar$, it should be natural to
 set the relaxation time as $\tau_\alpha^{}=\beta\hbar$.
Accordingly, in BCMD, $\mu_\alpha^{}$ is determined as
\begin{eqnarray}
 \mu_\alpha^{}  = \frac{\kappa_\alpha^{} \tau_\alpha^{}}{\gamma}.
 \label{eq2-12}
\end{eqnarray}
Note that with this setting the friction term,
 $\mu_\alpha^{} \gamma \dot{q}_\alpha^{}$,
 in the lhs of Eqs.(\ref{eq2-5}) and (\ref{eq2-9})
 does not depend on $\Delta t$.
As explained in the Appendix \ref{seca.a},
 $\beta\hbar$ corresponds to the time scale that free particle
 wavepacket persists at temperature $T=\frac{1}{k_{\rm B}^{}\beta}$.
Thus, BCMD maintains the shape of the free particle ring polymer,
 which is determined by the set of non-centroid variables,
 on the same time scale.
It seems that $\tau_\alpha^{}=\beta\hbar$ is a reasonable
 choice, at least, for the vibrational spectrum of OH molecule
 tested in Section \ref{sec3.2}.
%
In this case, Eq.(\ref{eq2-12}) combined with Eqs.(\ref{eq2-3})
 and (\ref{eq2-10}) becomes
 $\mu_\alpha^{}  = \frac{1}{2} m \omega_P^2 \lambda_\alpha^{}
 \Delta t (\beta\hbar)$.


In BCMD, one computes the correlation function
\begin{eqnarray}
 C_{AB}^{}(t) = 
 \left\langle \overline{A}(0) \overline{B}(t) \right\rangle
 \label{eq2-13}
\end{eqnarray}
 where $\overline{A}(0)$ and $\overline{B}(t)$
 is the bead average of the observable $X=A$ at time $0$
 and the observable $X=B$ at time $t$,
 respectively, which are given by
\begin{eqnarray}
 \overline{X}(t) = \frac{1}{P} \sum_{j=1}^P X(x_j^{}(t),p_j^{}(t)).
 \label{eq2-14}
\end{eqnarray}
In Sections \ref{sec2.3} and \ref{sec2.4}, it will be shown that
 Eq.(\ref{eq2-13})
 is an approximation of the Kubo-transformed
 correlation function, whose exact expression is
\begin{eqnarray}
 C_{AB}^{\rm kubo}(t) = \frac{ \frac{1}{\beta\hbar}
 \int_0^{\beta\hbar} d\tau {\rm Tr} \left( e^{-\beta\hat{H}}
 \hat{A}(0) \hat{B}(t+i\tau) \right) }
 { {\rm Tr} \left( e^{-\beta\hat{H}} \right) },
 \label{eq2-15}
\end{eqnarray}
where $\hat{H} = \frac{\hat{p}^2}{2m} + \hat{V}$
 is the Hamiltonian operator of the system.
Since BCMD does not disturb the thermal equilibria and
 is time reversible, the BCMD correlation
 function, Eq.(\ref{eq2-13}), is invariant under
 time origin shifts and time sign inversions.
Thus, $C_{AB}^{}(t)=C_{BA}^{}(t)$ is even function of $t$.
This is consistent with the exact correlation function,
 Eqs.(\ref{eq2-15}), which is real-valued and even function
 of $t$ with the same symmetry,
 $C_{AB}^{\rm kubo}(t)=C_{BA}^{\rm kubo}(t)$.
 \cite{craig2004quantum}

\subsection{\sffamily \large Short time correlations \label{sec2.3}}

Let us see how accurate $C_{xx}^{}(t)$ of Eq.(\ref{eq2-13}) is
 compared with $C_{xx}^{\rm kubo}(t)$ of Eq.(\ref{eq2-15})
 at short times.
Since $C_{xx}^{}(t)$ is an even function of $t$,
 the Taylor expansion of the position and velocity correlation
 functions near $t=0$ are described as
\begin{eqnarray}
 C_{xx}^{}(t) = C_{xx}^{(0)}
 + \frac{t^{2}}{2} C_{xx}^{(2)}
 + \frac{t^4}{24} C_{xx}^{(4)} + \cdots,
 \label{eq3-1}
\end{eqnarray}
and
\begin{eqnarray}
 C_{vv}^{}(t) = - \ddot{C}_{xx}^{}(t) =
 - C_{xx}^{(2)}
 - \frac{t^2}{2} C_{xx}^{(4)} - \cdots,
 \label{eq3-2}
\end{eqnarray}
respectively.
In Eq.(\ref{eq3-1}), the zeroth-order coefficient is
\begin{eqnarray}
 C_{xx}^{(0)}
 = \left\langle \overline{x}^2 \right\rangle
 = \left\langle {q}_1^{2} \right\rangle,
 \label{eq3-3}
\end{eqnarray}
the second-order coefficient is
\begin{eqnarray}
 C_{xx}^{(2)}
 = - \left\langle \dot{\overline{x}}^2 \right\rangle
 = - \left\langle \dot{q}_1^{2} \right\rangle
 = - \frac{1}{m^2} \left\langle p_1^{2} \right\rangle
 = - \frac{1}{m\beta},
 \label{eq3-4}
\end{eqnarray}
and the fourth-order coefficient is
\begin{eqnarray}
 && C_{xx}^{(4)}
 = - \left\langle \dot{\overline{x}} \dddot{\overline{x}} \right\rangle
 = - \frac{1}{m} \left\langle \dot{q}_1^{} \ddot{p}_1^{} \right\rangle
 = \left\langle \frac{\dot{q}_1^{}}{m} \frac{d}{dt}
 \frac{\partial \phi}{\partial q_1^{}} \right\rangle
 \nonumber \\ &&
 = \frac{1}{mP} \left\langle
 \sum_{j=1}^P \frac{\partial^2 V}{\partial x_j^{}}
 \dot{x}_j^{} \dot{q}_1^{} \right\rangle
 = \frac{1}{\beta m^2} \left\langle \frac{1}{P}
 \sum_{j=1}^P \frac{\partial^2 V}{\partial x_j^{2}} \right\rangle,
 \label{eq3-5}
\end{eqnarray}
where the principle of equipartition,
 $\frac{p_1^2}{m}=\frac{1}{\beta}$, has been used.
Eqs.(\ref{eq3-3}), (\ref{eq3-4}),
 and (\ref{eq3-5}), on the other hand, exactly match
 the zeroth, second, and fourth-order coefficients, respectively,
 in the Taylor expansion of the exact correlation function,
 $C_{xx}^{\rm kubo}(t)$.
 \cite{braams2006short}
Therefore, $C_{xx}^{}(t)$ and $C_{vv}^{}(t)$ are accurate up to
 the fifth order in $t$, and the third order in $t$, respectively.

\subsection{\sffamily \large Harmonic system \label{sec2.4}}

Let us consider the case of harmonic system with a potential energy
 function, $V(x) =\frac{K}{2} x^2$, where $K$ is the force constant.
In this case, Eq.(\ref{eq1-3}) becomes
\begin{eqnarray}
 \phi =
 \frac{K}{2P} \sum_{j=1}^P x_j^2
 = \frac{K}{2} \sum_{\alpha=1}^P q_\alpha^2.
 \label{eq4-1}
\end{eqnarray}
 where Eq.(\ref{eq1-6}) is used with the relation
 ${\bf U}^T = {\bf U}^{-1}$ in Eq.(\ref{eq1-5}).
For the CMD and RPMD equations of motion,
 Eqs.(\ref{eq1-9}) and (\ref{eq1-10}) become
\begin{eqnarray}
 m \ddot{q}_1^{} = - K q_1^{}, \ \ \mbox{and} \ \
 \mu_\alpha^{} \ddot{q}_\alpha^{} =
 - \kappa_\alpha^{\prime} q_\alpha^{},
 \label{eq4-2}
\end{eqnarray}
respectively, where
\begin{eqnarray}
 \kappa_\alpha^{\prime} \equiv m \omega_P^2 {\lambda}_\alpha^{} + K.
 \label{eq4-3}
\end{eqnarray}
For the BCMD equations of motion,
 Eqs.(\ref{eq2-9}) and (\ref{eq3-5}) become
\begin{eqnarray}
 m \ddot{q}_1^{} = - K q_1^{}, \ \ \mbox{and} \ \
 \mu_\alpha^{} \gamma \dot{q}_\alpha^{} = {\xi}_\alpha^{}
 - \kappa_\alpha^{\prime} {q}_\alpha^{}
 \label{eq4-4}
\end{eqnarray}
respectively.


In both cases of Eq.(\ref{eq4-2}) and (\ref{eq4-4}),
 the motion of the centroid variable is uncoupled
 from that of the non-centroid variables.
The time evolution of centroid variable is
\begin{eqnarray}
 {q}_1^{}(t) = {q}_1^{}(0) \cos(\omega t)
 + \dot{q}_1^{}(0) \omega^{-1} \sin(\omega t)
 \label{eq4-5}
\end{eqnarray}
where $\omega = \sqrt{\frac{K}{m}}$.
From Eq.(\ref{eq4-5}) the centroid position
 correlation function becomes
\begin{eqnarray}
 C_{xx}^{}(t) =
 \langle \overline{x}(0) \overline{x}(t) \rangle
 = \langle {q}_1^{}(0) {q}_1^{}(t) \rangle
 = \frac{\cos(\omega t)}{\beta m\omega^2},
 \label{eq4-6}
\end{eqnarray}
 where the principle of equipartition for harmonic oscillators,
 $\langle {q}_1^{2} \rangle = \frac{1}{\beta m\omega^2}$, 
 has been used.
From Eq.(\ref{eq4-6})
 the centroid velocity correlation function is
\begin{eqnarray}
 C_{vv}^{}(t)
 = - \ddot{C}_{xx}^{}(t)
 = \frac{\cos(\omega t)}{\beta m}.
 \label{eq4-7}
\end{eqnarray}
Eqs.(\ref{eq4-6}) and (\ref{eq4-7}) are exactly the same as
 $C_{xx}^{\rm kubo}(t)$ and $C_{vv}^{\rm kubo}(t)$, respectively,
 for the harmonic system.
 \cite{craig2004quantum}

\subsection{\sffamily \large Numerical algorithm \label{sec2.5}}

Generally, overdamped Langevin dynamics
 can be considered as short-time case of
 the hybrid Monte Carlo (HMC).
 \cite{kennedy1990theory}
The HMC algorithm
 \cite{duane1987hybrid,kennedy1990theory,mehlig1992hybrid,tuckerman2010statistical}
 is composed of the velocity randomization,
 the integration of Hamilton's equation of motion with a time reversible
 and area preserving algorithm, and the Metropolis acceptance/rejection.
The Metropolis acceptance/rejection can be skipped for the
 short-time case because the energy is conserved.
 \cite{kennedy1990theory}
An application of HMC to Eqs.(\ref{eq1-9}) and (\ref{eq1-10})
 corresponds to path integral hybrid Monte Carlo (PIMHC).
 \cite{tuckerman1993efficient,shiga2018path}
Thus the integration of the BCMD equation of motion can be
 built in a way similar to that of the PIHMC method
 based on the Reversible System Propagator Algorithm (RESPA),
 \cite{tuckerman2010statistical}
 which is time reversible and area preserving.
In fact, BCMD could be thought of as a special class of PIHMC
 applied only to the non-centroid variables
 where the trial move consists of a single step, and thus the
 velocity randomization of the non-centroid variables
 is applied each step.
%
This leads to a pseudo-code for the BCMD cycle as follows:
\begin{eqnarray}
 &&
 v_\alpha^{} \leftarrow v_\alpha^{\prime}\ (\mbox{randomized for}\ \alpha \ge 2)
 \nonumber \\ &&
 v_1^{} \leftarrow
 v_1^{} - \frac{\partial \phi}{\partial q_1^{}} \frac{\Delta t}{2m}, \ \ \ \
 v_\alpha^{} \leftarrow
 v_\alpha^{} - \frac{\partial \phi}{\partial q_\alpha^{}} \frac{\Delta t}{2\mu_\alpha^{}}
  \ \ (\alpha \ge 2)
 \nonumber \\ &&
 q_1^{} \leftarrow q_1^{} + v_1^{} \Delta t
 \nonumber \\ &&
 q_\alpha^{} \leftarrow
 q_\alpha^{} \cos( \omega_\alpha^{} \Delta t)
 + v_\alpha^{} \omega_\alpha^{-1} \sin( \omega_\alpha^{} \Delta t)
 \ \ (\alpha \ge 2),
 \nonumber \\ &&
 v_\alpha^{} \leftarrow
 v_\alpha^{} \cos( \omega_\alpha^{} \Delta t)
 - q_\alpha^{} \omega_\alpha^{} \sin( \omega_\alpha^{} \Delta t)
 \ \ (\alpha \ge 2)
 \nonumber \\ &&
 \mbox{calculate $\frac{\partial \phi}{\partial x_j^{}}$,
 obtain $\frac{\partial \phi}{\partial q_1^{}}$,
 $\frac{\partial \phi}{\partial q_\alpha^{}}$ from Eq.(\ref{eq1-12})}
 \nonumber \\ &&
 v_1^{} \leftarrow
 v_1^{} - \frac{\partial \phi}{\partial q_1^{}} \frac{\Delta t}{2m}, \ \ \ \
 v_\alpha^{} \leftarrow
 v_\alpha^{} - \frac{\partial \phi}{\partial q_\alpha^{}} \frac{\Delta t}{2\mu_\alpha^{}}
 \ \ (\alpha \ge 2)
 \label{eq5-1}
\end{eqnarray}
where
\begin{eqnarray}
 \omega_\alpha^{}
 = \sqrt{\frac{m\lambda_\alpha^{}}{\mu_\alpha^{}}} \omega_P^{}.
 \label{eq5-2}
\end{eqnarray}
In Eq.(\ref{eq5-1}), line 1 corresponds to the
 velocity randomization, while the rest corresponds to
 the time integration of Eqs.(\ref{eq1-9}) and (\ref{eq1-10}).
To ensure time reversibility,
 the time integration for the force contributions
 from the first and second terms in the rhs of Eq.(\ref{eq1-2})
 are divided into lines (4,5) and lines (2,3,7), respectively,
%
 and the analytic expression for the harmonic oscillator is
 adopted for the former, which corresponds to taking
 the number of steps infinite for
 the reference forces in the RESPA technique.
%
The force for the second term ($\phi$)
 is calculated at line 6.
The velocity update at Line 7 is necessary not only
 for the centroids but also for the non-centroids
 in order to assess the energy conservation of $E^\prime(t)$,
 see Section \ref{sec2.7}.


%
BCMD gives rise to the canonical ensemble as in PIHMC,
 but it has access to the time dependence via Kubo-transformed
 correlation function in contrast to PIHMC\@.
A notable difference of the BCMD algorithm from the PIHMC algorithm
 is that the Metropolis acceptance/rejection
 step can be omitted.
This is because in BCMD a small step size
 $\Delta t$ is chosen such that the energy
 is conserved within a single step.
In other words, $H$ defined in Eq.(\ref{eq1-7})
 is almost constant when Eq.(\ref{eq1-8}) is solved
 for a single step excluding the velocity randomization part
 (as in the case of PIHMC algorithm),
 which corresponds to Lines 2--7 in Eq.(\ref{eq5-1}).
This is in contrast to the case of PIHMC where a large
 step size $\Delta t$ is used such that $H$ is no longer conserved,
 and Eq.(\ref{eq5-1}) is regarded as a trial move.
In PIHMC, a set of the cycle(s) of Eq.(\ref{eq5-1}) is accepted with
 the probability $P={\min(1,e^{-\beta\Delta H})}$ where $\Delta H$
 is the change in $H$ upon the trial move.
In BCMD, meanwhile, $P\approx 1$ if $H \ll \beta^{-1}$,
 and thus the trial move is always accepted.
%
%
The BCMD algorithm is similar to the RPMD algorithm, except
 for the velocity randomization and the values for $\mu_\alpha^{}$.


The ``partially adiabatic'' CMD (PACMD) using a non-large
 adiabaticity parameter, $\gamma_{\rm cmd}^{}$,
 would have an aspect similar to BCMD in that the adiabatic separation
 between centroid and noncentroid variables becomes incomplete.
However, PACMD is difficult to establish without ambiguity
 in setting up the non-centroid masses and the system-thermostat
 coupling.
Thus the results of PACMD should strongly depend on such setups.
In contrast, BCMD is uniquely defined
 by Eqs.(\ref{eq1-9}) and (\ref{eq2-5})
 [Eqs.(\ref{eq6-4}) and (\ref{eq6-7}) for the multidimensional case],
 and the degree of nonadiabaticity is determined unambiguously.


%
%
%
The simple algorithm Eq.(\ref{eq5-1}) works correctly
 as long as the step size is small enough.
It is required that the energy is conserved, which
 could be checked numerically by monitoring the energy
 conservation of $E^\prime(t)$, as mentioned in Section \ref{sec2.7}.
It may also be useful to confirm that the non-centroid MSD
 in the absence of physical potential, Eq.(\ref{eq2-11}),
 is reproduced with the step size chosen.
A modern algorithmic development of the Langevin equation,
 \cite{liu2016simple,zhang2017unified,li2017stationary}
 especially those beyond the simple ``pxpT'' algorithm,
 \cite{li2017stationary}
 may help making it more efficient and accurate by enabling
 the use of increased step sizes.




\subsection{\sffamily \large Multidimensional extension \label{sec2.6}}

Let us consider a system composed of $N$ atoms in 3-dimensional space
 with Cartesian coordinates $({\bf R}_1^{},\cdots,{\bf R}_N^{})$, 
 and the potential energy function $V({\bf R}_1^{},\cdots,{\bf R}_N^{})$.
Following Section \ref{sec2.1}, the cyclic path of the $I$-th atom,
 ${\bf R}_I^{} = \left( {\bf R}_I^{(1)},
 \cdots, {\bf R}_I^{(P)} \right)^T$,
 can be represented in normal modes
 ${\bf Q}_I^{} = \left( {\bf Q}_I^{(1)},
 \cdots, {\bf Q}_I^{(P)} \right)^T$, using
 the linear transformation
\begin{eqnarray}
 {\bf Q}_I^{} = \frac{1}{\sqrt{P}} {\bf U}^T {\bf R}_I^{}.
 \label{eq6-1}
\end{eqnarray}
With the set $\{{\bf Q}\}$ and its momentum conjugate $\{{\bf P}\}$,
 the Hamiltonian is constructed as
\begin{eqnarray}
 H(\{{\bf P},{\bf Q}\})
 =  \sum_{I=1}^N \left( \frac{1}{2}{\bf P}_I^T
 \boldsymbol{\mu}_I^{-1} {\bf P}_I^{}
 + \frac{1}{2}
 M_I^{} \omega_P^2 {\bf Q}_I^T \boldsymbol{\lambda} {\bf Q}_I^{} \right)
 + \phi(\{{\bf Q}\}),
 \label{eq6-2}
\end{eqnarray}
 where $M_I^{}$ is the physical
 atomic mass of the $I$-th atom, $\boldsymbol{\mu}_I^{}$ is
 a diagonal matrix whose elements are
 the fictitious normal mode masses,
 and
\begin{eqnarray}
 \phi(\{{\bf Q}\}) \equiv \frac{1}{P}
 \sum_{j=1}^{P} V\left({\bf R}_1^{(j)}
 ({\bf Q}_1^{}),\cdots,{\bf R}_N^{(j)}({\bf Q}_N^{})\right)
 \label{eq6-3}
\end{eqnarray}
 is the average physical potential.
The canonical equation of motion derived
 from Eq.(\ref{eq6-2}) is as follows.
For the centroid variables,
\begin{eqnarray}
 \dot{\bf Q}_{I}^{(1)}
 = \frac{{\bf P}_{I}^{(1)}}{M_I^{}}
 = {\bf V}_{I}^{(1)}, \ \ \ \
 \dot{\bf P}_{I}^{(1)} =
 - \frac{\partial \phi}{\partial {\bf Q}_{I}^{(1)}},
 \label{eq6-4}
\end{eqnarray}
and for the non-centroid variables $\alpha \ge 2$,
\begin{eqnarray}
 \dot{\bf Q}_{I}^{(\alpha)} =
 \frac{{\bf P}_{I}^{(\alpha)}}{\mu_{I}^{(\alpha)}}
 = {\bf V}_{I}^{(\alpha)}, \ \ \ \
 \dot{\bf P}_{I}^{(\alpha)} = {\bf F}_I^{(\alpha)},
 \label{eq6-5}
\end{eqnarray}
with
\begin{eqnarray}
 {\bf F}_I^{(\alpha)} =
 - M_I^{} \omega_P^2 {\lambda}^{(\alpha)} {\bf Q}_{I}^{(\alpha)}
 - \frac{\partial \phi}{\partial {\bf Q}_{I}^{(\alpha)}}.
 \label{eq6-6}
\end{eqnarray}
In BCMD, the non-centroid velocities
 are randomized at each step (${\bf V}_{I}^{(\alpha)\prime}$).
Then, following Section \ref{sec2.2},
 one obtains the Brownian equation
\begin{eqnarray}
 {\mu}_I^{(\alpha)} \gamma \dot{\bf Q}_I^{(\alpha)}(t)
 = \boldsymbol{\xi}_I^{(\alpha)}(t) + {\bf F}_I^{(\alpha)}(t),
 \label{eq6-7}
\end{eqnarray}
where
\begin{eqnarray}
 \boldsymbol{\xi}_I^{(\alpha)}
 = \sqrt{\frac{2\mu_I^{(\alpha)} \gamma}{\beta}}\boldsymbol{\zeta}(t)
 \label{eq6-8}
\end{eqnarray}
is the random force.
Eqs.(\ref{eq6-7}) and (\ref{eq6-8}) are the working
 BCMD equation of motion for multidimension systems.
The masses are set to
 $\mu_I^{(1)} = M_I^{}$ for the centroids, and
\begin{eqnarray}
 \mu_I^{(\alpha)} = \frac{1}{2} M_I^{}
 \omega_P^2 \lambda^{(\alpha)} ({\Delta t}) \tau_\alpha^{},
 \label{eq6-9}
\end{eqnarray}
 with $\tau_\alpha^{}=\beta\hbar$
 for the non-centroids ($\alpha \ge 2$),
 which is obtained by combining
 Eq.(\ref{eq2-12}) with Eqs.(\ref{eq2-3}) and (\ref{eq2-10}).

\subsection{\sffamily \large Conservation laws \label{sec2.7}}

To calculate the vibrational spectrum, it is useful to remove
 the total momentum.
In the case of gaseous systems in the free boundary condition,
 it is also useful to remove the full angular momentum.
If the potential $V$ has the translation symmetry,
 BCMD conserves the total centroid momentum,
 ${\bf P}_{\rm cent}^{} \equiv
 \sum_{I=1}^N M_I^{} \dot{\bf Q}_{I}^{(1)}$,
 because
\begin{eqnarray}
 \dot{\bf P}_{\rm cent}^{}
 = \sum_{I=1}^N M_I^{} \ddot{\bf Q}_{I}^{(1)}
 = - \sum_{I=1}^N
 \frac{\partial \phi}{\partial {\bf Q}_{I}^{(1)}}
 = - \frac{1}{P} \sum_{k=1}^P \left( \sum_{I=1}^N
 \frac{\partial V}{\partial {\bf R}_{I}^{(k)}} \right) = 0.
 \label{eq7-1}
\end{eqnarray}
Thus, ${\bf P}_{\rm cent}^{} = 0$ is satisfied
 by setting at the initial step,
\begin{eqnarray}
 \dot{\bf Q}_{I}^{(1)} \leftarrow
 \dot{\bf Q}_{I}^{(1)}
 - \frac{\sum_{J=1}^N M_J^{} \dot{\bf Q}_{J}^{(1)}}{\sum_{J=1}^N M_J^{}}.
 \label{eq7-2}
\end{eqnarray}
On the other hand, BCMD does not conserve by itself
 the total angular momentum of the centroid variables,
 ${\bf L}_{\rm cent}^{}
 \equiv \sum_{I=1}^N \left( {\bf Q}_{I}^{(1)} \times
 M_I^{} \dot{\bf Q}_{I}^{(1)}\right)$.
However, ${\bf L}_{\rm cent}^{}=0$ can be imposed in BCMD
 using the rotational correction technique,
 as done in CMD and RPMD as well. \cite{witt2009applicability}
This technique removes at each step the rotational component of
 the centroid velocities by
\begin{eqnarray}
 \dot{\bf Q}_{I}^{(1)} \leftarrow \dot{\bf Q}_{I}^{(1)}
 - \left( {\bf I}_{\rm cent}^{-1} {\bf L}_{\rm cent}^{} \right)
 \times {\bf Q}_{I}^{(1)},
 \label{eq7-3}
\end{eqnarray}
and the rotational component of the centroid forces by
\begin{eqnarray}
 {\bf F}_{I}^{(1)} \leftarrow {\bf F}_{I}^{(1)}
 - \left( {\bf I}_{\rm cent}^{-1} {\bf N}_{\rm cent}^{} \right)
 \times {\bf Q}_{I}^{(1)},
 \label{eq7-4}
\end{eqnarray}
 where ${\bf I}_{\rm cent}^{}$ is a $3\times 3$ matrix for
 the centroid moment of inertia, and
 ${\bf N}_{\rm cent}^{}$ is a $3N$-dimensional vector for
 the centroid torque.


The energy conservation of the classical
 isomorph is useful in determining a reasonable
 step size, $\Delta t$.
In the BCMD algorithm,
 the energy conservation for $H(\{{\bf P},{\bf Q}\})$
 defined in Eq.(\ref{eq6-2}) is kept in the time
 integration part, but is broken in the
 velocity randomization part.
For this reason, one can introduce a modified energy for BCMD as
\begin{eqnarray}
 E^\prime(t) = H(\{{\bf P},{\bf Q}\})
 - \sum_{s<t}^{} \sum_{I=1}^N \sum_{\alpha=2}^P \frac{\mu_I^{(\alpha)}}{2}
 \left( \left| {\bf V}_I^{(\alpha)\prime}(s) \right|^2
 -\left|{\bf V}_I^{(\alpha)}(s)\right|^2 \right),
 \label{eq7-5}
\end{eqnarray}
 where the second term in the rhs corrects the jumps in
 the kinetic energy due to the velocity randomization.
$E^\prime(t)$ is conserved for small step size $\Delta t$
 in the absence of the rotational correction.
When the rotational correction is applied,
 $E^\prime(t)$ is not strictly conserved.
However, the energy shift due to the rotational correction
 is small, as was experienced in CMD and RPMD as well.
Thus the conservation error of $E^\prime(t)$ is acceptable within
 tens of picoseconds of BCMD trajectories required in
 the computation of vibrational spectra.

\section{\sffamily \large CALCULATIONS \label{sec3}}

All the calculations in this study were performed using
 the in-house version of the \texttt{PIMD} software,
 \cite{shiga2020pimd,shiga2001unified,shiga2018path}
 with the BCMD method newly implemented.

\subsection{\sffamily \large One dimensional models \label{sec3.1}}

The BCMD method was first tested on one-dimensional
 models, as done in previous studies of the CMD and RPMD methods.
 \cite{jang1999derivation,craig2004quantum,hone2006comparative,perez2009comparative}
Figures \ref{fig1} and \ref{fig2} show the Kubo-transformed
 correlation functions $C_{xx}^{}(t)$ for
 mildly anharmonic and quartic potentials, respectively.
%
Figure \ref{fig3} shows the function $C_{aa}^{}(t)$ for a
 harmonic potential, where $a=x^2$.
%
The BCMD results were compared with the results
 of MD, CMD, RPMD, and the exact solution.
%
The CMD and RPMD results are consistent with earlier works
 such as Figures 1--2 in Reference
 \cite{craig2004quantum}, Figures 7--8 in
 Reference \cite{perez2009comparative},
 and Figure 1 of Reference \cite{liu2011two}.
%
Note in Figure \ref{fig3} of this paper
 that the correlation functions
 where nonlinear operators are involved
 are not necessarily exact in BCMD as well as CMD and (T)RPMD
 even for a harmonic system.


In general, the performance of BCMD
 shown in Figures \ref{fig1}--\ref{fig3} of this paper
 is similar to those of CMD and RPMD\@.
However, there is a slight difference in the low temperatures case of
 quartic potential shown in the bottom panel of Figure \ref{fig2}.
Here it can be seen that the BCMD captures the amplitude of
 long-time vibrations better than RPMD, but worse than CMD\@.
The stronger the anharmonicity and the lower the temperature,
 the faster the deviation from the exact solution will be.
BCMD, like CMD and RPMD, cannot describe
 long-time quantum interference in
 anharmonic oscillations, as expected from theory.
This means that this method is more suitable for condensed
 phase systems where the interference effects are expected
 to be rapidly quenched.

\subsection{\sffamily \large OH molecule \label{sec3.2}}

Next, the BCMD method was tested on the vibrational
 spectra for the harmonic and Morse models of gaseous OH molecule.
These models were used in previous studies to test the CMD, RPMD
 and TRPMD methods.
 \cite{witt2009applicability,rossi2014remove}
The harmonic model is not a quadratic function
 with respect to the Cartesian coordinates of the O and H atoms, but a
 quadratic function with respect to the shift of the beadwise OH distance in
 three dimensional space from the equilibrium distance.
For this reason, BCMD, CMD, and RPMD do not
 provide the exact results for this model.


The BCMD, CMD, and RPMD simulations were performed at temperatures
 100, 200, 300 and 600 K with $P=96$, 48, 32 and 16 beads, respectively.
%
Starting from the equilibrated structures of the PIMD
 simulations, 100--300 trajectories of 2.5 ps long
 were computed for each case.
For CMD, the adiabaticity parameter was set
 to $\gamma_{\rm cmd}^{}=10$\@.
The rotational correction was applied
 to fix the molecular orientation.
The MD simulations were performed in the same way
 as above, but with a single bead, $P=1$.


Previous studies pointed out that there are
 chain resonance and curvature problems
 in the RPMD and CMD vibrational spectra, respectively.
When $P$ is large, the chain resonance is found at the RPMD
 normal mode frequencies,
 $\omega_n^{}=\frac{2n\pi}{\beta\hbar}$
 where $n$ is a natural number.
At 300 K, $\omega_n^{} = 1310\times n$ cm$^{-1}$,
 are within the frequency range of physical vibrations
 (0--5000 cm$^{-1}$).
As the temperature decreases, the amount of resonant modes
 in this region increases, making the spurious peak splittings
 become more noticable.
This is shown in the bottom right panel of Figures
 \ref{fig4} and \ref{fig5}.
%
TRPMD diminishes the chain resonance, but it does not completely
 remove it even with the optimal damping parameter.
%
On the other hand, the curvature problem arises from the spurious
 coupling between the rotation and vibration of the system.
The coupling is amplified by the adiabaticity between the centroid
 and non-centroid variables which is assumed in the CMD equation
 of motion.
This leads to an unnatural red shift and broadening of the spectra
 as the temperature decreases, as shown in the bottom-left panel
 of Figures \ref{fig4} and \ref{fig5}.


The results of BCMD are shown in the top-left panels
 of Figures \ref{fig4} and \ref{fig5}.
The peak positions from Figure \ref{fig5} were plotted
 in the left panel of Figure \ref{fig6}.
An advantage of BCMD is in the complete absence of peak
 splitting due to spurious resonances, which is often a
 fundamental requirement in vibrational spectroscopy.
On the other hand, the peak shift
 due to the curvature problem of BCMD does not seem
 to be as severe as CMD\@.
The reason is that BCMD
 breaks the adiabaticity between the centroid and non-centroid
 variables, and thus weakens the spurious coupling
 between rotation and vibration.
Still, in this case, the peak shift of BCMD cannot
 be ignored, especially at temperatures below 200 K in this case.
The BCMD peak positions reasonably matches
 the exact frequency of the $(v,r) = (0,0) \rightarrow (1,0)$
 transition in both the harmonic and Morse models.
For the Morse model, the peak position in BCMD (top-left panel)
 shows a redshift relative to that of MD (top-right panel),
 improving the agreement with the exact result.
This shows that BCMD properly considers the nuclear
 quantum effects of anharmonic vibrations.


The peak width for the Morse model are shown in the right panel of Figure
 \ref{fig6}.
Here the full-width at half minimum (FWHM) values were evaluated by
 fitting the spectrum to the Lorenzian function.
This result shows the limitation of BCMD that unphysical broadening
 of the spectrum becomes noticeable at low temperatures,
 especially below 200 K in this case.
CMD and TRPMD also show more or less  unphysical broadening
 at low temperatures as well.


Further tests on the vibrational spectra of
 methane molecule is shown in Appendix B\@.
%
As a general trend, the performance of BCMD shown
 in Figure \ref{fig13} and \ref{fig14} was similar
 to those seen in Figures \ref{fig4} and \ref{fig5}.
The unphysical peak shift and broadening were found to decrease
 for the vibrational modes with lower frequency.
They are generally smaller for the HOH and HCH bending modes
 than in the OH and CH stretching modes,
 and they are smaller in the D isotopologues
 than in the H isotopologues.

\subsection{\sffamily \large Gaseous water \label{sec3.3}}

The BCMD, CMD, RPMD, and MD simulations of gaseous water molecule
 and its isotopologues
 were carried out at 300 K using the q-TIP4P/F model.
 \cite{habershon2009competing}
The simulation settings were the same as those described
 in Section \ref{sec3.2}.
In Figure \ref{fig7}, the results are compared
 to the solution of vibrational Schr\"odinger
 equation based on the full
 vibrational configuration interaction (VCI) method
 \cite{yagi2000direct}
 using the SINDO code,
 \cite{SINDO}
 with a 3-mode representation and 11 grid points per mode.
Since the q-TIP4P/F potential is
 quartic with respect to the OH bonds,
 the exact peak frequencies in the spectra are
 red-shifted from the harmonic frequencies.


In all cases studied
 for H$_2^{}$O, D$_2^{}$O, HDO and HTO molecules,
 the BCMD vibrational spectra were
 significantly improved compared to
 the MD vibrational spectra.
This indicates that the nuclear quantum effects
 on the vibrational spectra
 were properly taken into account in the BCMD method.
Assuming that the VCI frequencies are correct,
 the mean absolute error and the maximum absolute error of the BCMD
 peaks were 22 and 41 cm$^{-1}$, respectively.
These values are lower than the mean absolute error and
 the maximum absolute error of
 the CMD peaks (44 and 129 cm$^{-1}$, respectively),
 the RPMD peaks (52 and 156 cm$^{-1}$, respectively),
 the MD peaks (77 and 151 cm$^{-1}$, respectively),
 and the harmonic frequencies (85 and 166 cm$^{-1}$, respectively).
In this respect, the BCMD method outperforms the CMD and RPMD methods
 in these cases.
%
However, the spectral broadening of the BCMD method remains a problem
 when dealing with vibrational spectra of the gas phase molecules
 at the low temperature, and it seems that PILD outperforms BCMD
 in this sense.
 \cite{liu2016path}

\subsection{\sffamily \large Liquid water \label{sec3.4}}

As a typical example of condensed phase systems, the BCMD method was
 tested on the vibrational spectra of liquid water and its isotopologues
 at a temperature of 300 K\@.
TIP3P and modified TIP3P (mTIP3P) models were used for comparison
 with previous studies on MD and CMD simulations.
 \cite{ivanov2010communications}
%
In addition, the q-TIP4P/F model was used for comparison
 with previous studies on MD and (T)RPMD simulations.
 \cite{rossi2014remove}
%
The BCMD simulations of liquid H$_2^{}$O were performed
 using $P=32$ beads, for a system
 of 256 H$_2^{}$O molecules contained in a cubic box with a side
 length of 19.7 \AA.
Ewald sum technique was used to compute the electrostatic
 interaction under the periodic boundary condition.
Starting from the equilibrated structures of
 the PIMD simulations, 100 independent BCMD trajectories were
 run for 2.5 ps each with a step size of $\Delta t = 0.25$ fs.
The same procedure was repeated for the BCMD simulations of
 liquid D$_2^{}$O and HDO\@.
The intensity of the vibrational spectra was computed
 by the formula
\begin{eqnarray}
 \alpha(\omega) \propto \beta\omega^2 \int_{-\infty}^\infty
 C_{\bf MM}^{}(t) \exp(-{\rm i}\omega t) dt
 \label{eq8-1}
\end{eqnarray}
 where $C_{\bf MM}^{}(t)$ is the Kubo-transformed dipole
 autocorrelation function.
In this case, the dipole trajectories were computed by
\begin{eqnarray}
 {\bf M}^{(j)}(t) = \sum_{I=1}^N \rho_I^{} {\bf R}_I^{(j)}(t)
 \label{eq8-2}
\end{eqnarray}
 where ${\bf R}_I^{}(t)$ and $\rho_I^{}$
 are the Cartesian coordinates (in the expanded space
 of periodic boundary) and the TIP3P charge,
 respectively, of the $I$-th atom.


Figure \ref{fig8} shows the BCMD results compared with
 the MD, CMD, and (T)RPMD results.
Here, the spectral peaks correspond respectively to
 intermolecular vibrations (0-1000 cm$^{-1}$),
 DOD bending (1200--1300 cm$^{-1}$),
 HOD bending (1500--1600 cm$^{-1}$),
 HOH bending (1700--1800 cm$^{-1}$),
 the OD stretching (2200--2500 cm$^{-1}$), and
 the OH stretching (2900--3400 cm$^{-1}$).
It was pointed out that CMD shows a significant
 redshift in the spectra compared to MD, partly
 due to the curvature problem. \cite{ivanov2010communications}
It is expected that the correct peak positions
 is somewhere between those of MD and CMD,
 when the physical redshifts due to the nuclear quantum
 effects were taken into account.
In this respect, the results of BCMD are indeed convincing,
 since the peak positions are found systematically between
 those of MD and CMD\@.
Thus it seems that BCMD is capable of improving CMD
 with respect to the vibrational spectra of liquid water.
%
Comparing the results of BCMD and TRPMD, the peaks of
 the OH stretching and HOH
 bending were found similar to each other,
 but the peak of intermolecular vibrations in BCMD were
 found at higher frequency than that in TRPMD\@.
 \cite{rossi2014remove}

\subsection{\sffamily \large Protonated water: Ab initio BCMD \label{sec3.5}}

The final test is on the applicability of ab initio BCMD simulations
 where the electronic structure calculations are performed on the fly.
Here the infrared spectra of the protonated water dimer (Zundel ion),
 H$_5^{}$O$_2^{+}$, and its isotopologue, D$_5^{}$O$_2^{+}$,
 were taken as an example.
In this system, the anharmonicity in vibration due to quantum
 and thermal fluctuations is important.
 \cite{tachikawa2005geometrical}
%


Electronic structure calculations were
 performed using the \texttt{SMASH} code
 \cite{ishimura2017smash}
 which was built into the \texttt{PIMD} software.
 \cite{ruiz2016hierarchical}
Potential energy ($V$) and force ($-\nabla V$)
 were calculated by
 the MP2 method with the 6-31G(d) basis set.
 \cite{ishimura2006new,ishimura2007new}
The infrared spectrum was calculated from
 the dipole correlation function, Eq.(\ref{eq8-1}).
In this case the dipole trajectories were computed
 by the Hartree-Fock (HF) theory
 using the 6-31G(d) basis set,
\begin{eqnarray}
 {\bf M}^{(j)}(t) =
 \sum_{I=1}^N Z_I^{} {\bf R}_I^{(j)}(t)
 - \langle \phi^{(j)}(t) |
 \sum_{i=1}^n e \hat{\bf r}_i^{} |
 \phi^{(j)}(t) \rangle,
 \label{eq8-3}
\end{eqnarray}
 where $e$, $n$, $\hat{\bf r}_i^{}$, $Z_I^{}$, and $\phi^{(j)}(t)$
 represent the elementary charge, the number of electrons,
 the position operator of the $i$-th electron,
 the nuclear charge of the $I$-th nucleus, and the HF
 wavefunction of the $j$-th bead at time $t$, respectively.
Ab initio PIMD and BCMD simulations with $P=24$ beads
 were performed at the temperature 300 K\@.
After the equilibration with the ab initio PIMD simulation,
 ab initio BCMD simulations were run for
 50 trajectories, each with the length of 2.5 ps.
The rotational correction was applied
 to fix the orientation of the system.
The step size was set to $\Delta t = 0.25$ fs.
The ab initio MD simulations were performed in the same way
 but using a single bead, at the temperature 300 K\@.


The results are shown in Figure \ref{fig9}.
It can be seen that the infrared spectra calculated
 from ab initio BCMD are overall in good agreement
 with the experimental results \cite{guasco2011unraveling}
 for both H$_5^{}$O$_2^{+}$ and D$_5^{}$O$_2^{+}$.
%
 The spectra of H$_5^{}$O$_2^{+}$ is also in reasonable
 agreement with the optimally-damped TRPMD calculations
 based on ab initio-based force field.
 \cite{rossi2014remove}
%
For the OD and OH stretching peaks near 2600 cm$^{-1}$
 and 3600 cm$^{-1}$, respectively,
 the BCMD peak positions are redshifted compared
 to the MD peak positions,
 and are closer to the experimental peak positions.
Also for the DOD and HOH bending peaks near 1300 cm$^{-1}$
 and 1800 cm$^{-1}$, respectively,
 the BCMD peak positions are redshifted compared
 to the MD peak positions,
 and are closer to the experimental peak positions.
For the peaks of the shared hydrogen vibrational modes
 of D$_5^{}$O$_2^{+}$ and H$_5^{}$O$_2^{+}$
 near 800 cm$^{-1}$ and 1100 cm$^{-1}$, respectively,
 the BCMD peak positions are blueshifted compared
 to the MD peak positions, and are higher than
 the experimental peak positions.
However, because this peak is temperature sensitive,
 it may not be fair to directly compare theory and experiments
 with different temperature settings.
In fact, it was experimentally observed that
 the peak in the Ar-tagged H$_5^{}$O$_2^{+}$
 is blueshifted compared to the Ne-tagged H$_5^{}$O$_2^{+}$.
 \cite{guasco2011unraveling}
Further research should be necessary to verify this issue.

\section{\sffamily \large CONCLUSIVE REMARKS \label{sec4}}

In this paper, the BCMD method was proposed as a new approximation
 of quantum vibrational dynamics.
BCMD is similar to CMD and RPMD in its accuracy in the short-time
 correlation, in the classical/high temperature limits,
 and in the harmonic oscillator systems.
However, in the computation of vibrational spectra, BCMD eliminates
 the chain resonance problem of RPMD, and it alleviates the
 curvature problem of CMD\@.
The BCMD causes spectral broadening at low temperatures,
 which one should be most cautious of.
BCMD requires about the same computational effort as RPMD and CMD\@.
The ab initio BCMD simulation can be run in parallel computing, and
 should be useful for a wide range of applications.


\section{\sffamily \large APPENDIX}
\subsection*{\sffamily \large A: Wavepacket propagation \label{seca.a}}

Following the textbook by Schatz and Ratner, \cite{schatz2002quantum}
 a free particle wavepacket is expressed by the superposition
 of plane waves
\begin{eqnarray}
 \Psi(x,t)
 = N_0^{} \int_{-\infty}^\infty c(p)
 \exp\left( - \frac{p^2}{2m} \frac{{\rm i}t}{\hbar}
 +\frac{{\rm i}px}{\hbar} \right) dp,
 \label{eq9-1}
\end{eqnarray}
where $N_0^{}$ is a normalization constant, and $c(p)$ is a function
 describing the distribution of momentum, $p$, in the wavepacket.
Assuming the Maxwell-Boltzmann distribution at temperature
 $T=\frac{1}{k_{\rm B}^{}\beta}$, $c(p)=\sqrt{w(p)}$ and
%
\begin{eqnarray}
 w(p) = \sqrt{\frac{\beta}{2\pi m}} e^{- \beta \frac{p^2}{2m} },
 \label{eq9-2}
\end{eqnarray}
the probability distribution of the wavepacket is
\begin{eqnarray}
 |\Psi(x,t)|^2 = 
 \sqrt{ \frac{\pi}{a(t)} }
 \exp\left( - a(t) x^2 \right),
 \label{eq9-3}
\end{eqnarray}
 where
 $a(t)=\frac{ 2 \beta m}{(2t)^2+(\hbar\beta)^2}$
 and
 $N_0^{} =\frac{1}{\sqrt{2\pi\hbar}}$.
Thus, the wavepacket starts changing its shape
 after about $t = \frac{\beta\hbar}{2}$,
 and it is broadened by $\sqrt{5} \approx 2.2$ times
 at $t =\beta\hbar$.


%
%
%
%

\subsection*{\sffamily \large B: Gaseous methane \label{seca.c}}

Figures \ref{fig13} and \ref{fig14} show the infrared spectra
 of gaseous CH$_4^{}$ and CD$_4^{}$ molecules,
 respectively, of a harmonic force field used in
 Reference \cite{witt2009applicability}.
The potential energy function of this model is given by
\begin{eqnarray}
 V = \sum_{j\in{\rm CH}}^4 \frac{K_r^{}}{2} (r_j^{} - R^{})^2
 + \sum_{j\in{\rm HCH}}^6 \frac{K_\theta^{}}{2} (\theta_j^{} - \Theta)^2
 \label{eq9-5}
\end{eqnarray}
 where $r_j^{}$ is the $j$-th CH bond length,
 $\theta_j^{}$ is the $j$-th HCH bond angle, $R = 2.0598$ bohr,
 $K_r^{} = 0.3035$ hartree$\cdot$bohr$^{-2}$
 $\Theta = 107.8$ deg.
 $K_\theta^{} = 3.1068 \times 10^{-5}$ hartree$\cdot$deg.$^{-2}$.
The BCMD, CMD, RPMD, and MD simulations were carried out in the same
 manner as done for gaseous OH molecule in Section \ref{sec3.2}.

\section*{\sffamily \large ACKNOWLEDGMENTS}

The author acknowledges the financial support from
 ``Hydrogenomics'' of Grant-in-Aid for Scientific Research
  on Innovative Areas, MEXT, Japan,
  JSPS KAKENHI (18H05519, 21H01603), and
  the Supercomputer Fugaku (Fugaku Battery \& Fuel Cell Project).
The computations were conducted using the supercomputer facilities
  at Japan Atomic Energy Agency (JAEA).
%
The author thanks Dr. Bo Thomsen in JAEA and Dr. Kiyoshi Yagi
 in RIKEN for help in the VCI calculations.
%
The author is grateful to Noriko and Yohei Iwata for help reproducing
 figures from literatures.

\section*{\sffamily \large DATA AVAILABILITY}

The data that support the findings of this study are available
  within the article.

\section*{\sffamily \large CONFLICTS OF INTEREST}

There are no conflicts to declare.



\clearpage

\section{\sffamily \large FIGURE CAPTIONS}

\begin{figure}[htbp]
\begin{center}
\includegraphics[width=0.80\linewidth]{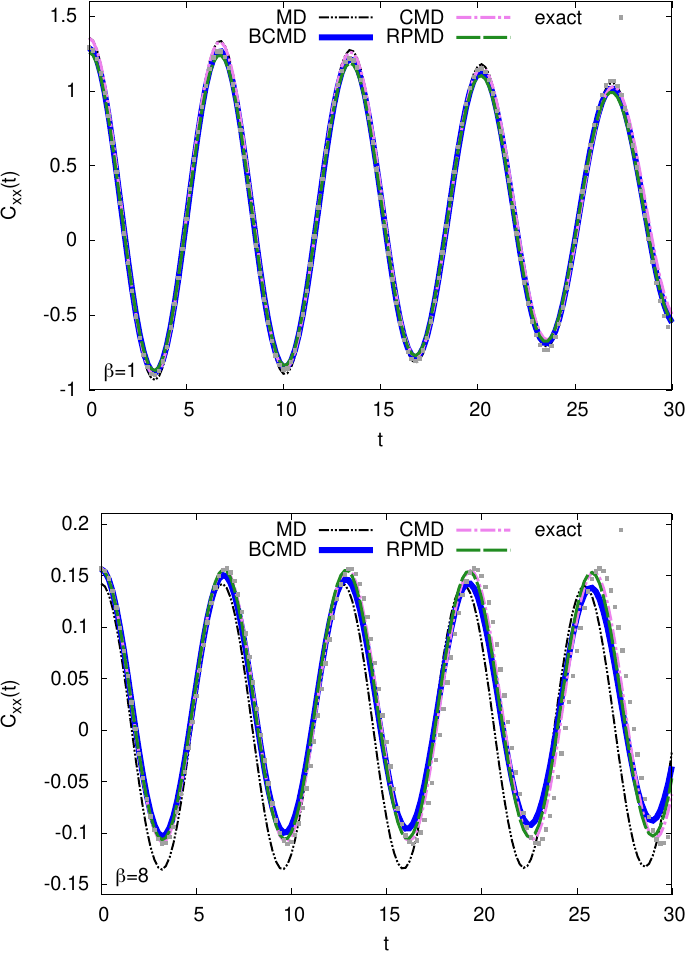}
\end{center}
\caption{Shiga, submitted to JCC.}
\label{fig1}
\end{figure}

\begin{figure}[htbp]
\begin{center}
\includegraphics[width=0.80\linewidth]{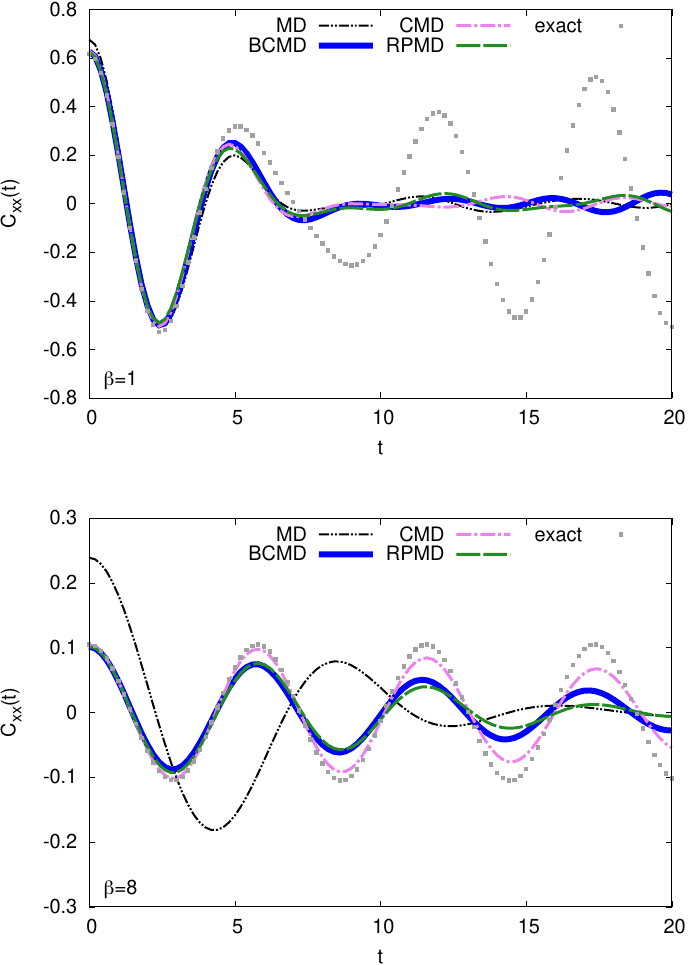}
\end{center}
\caption{Shiga, submitted to JCC.}
\label{fig2}
\end{figure}

\begin{figure}[htbp]
\begin{center}
\includegraphics[width=0.80\linewidth]{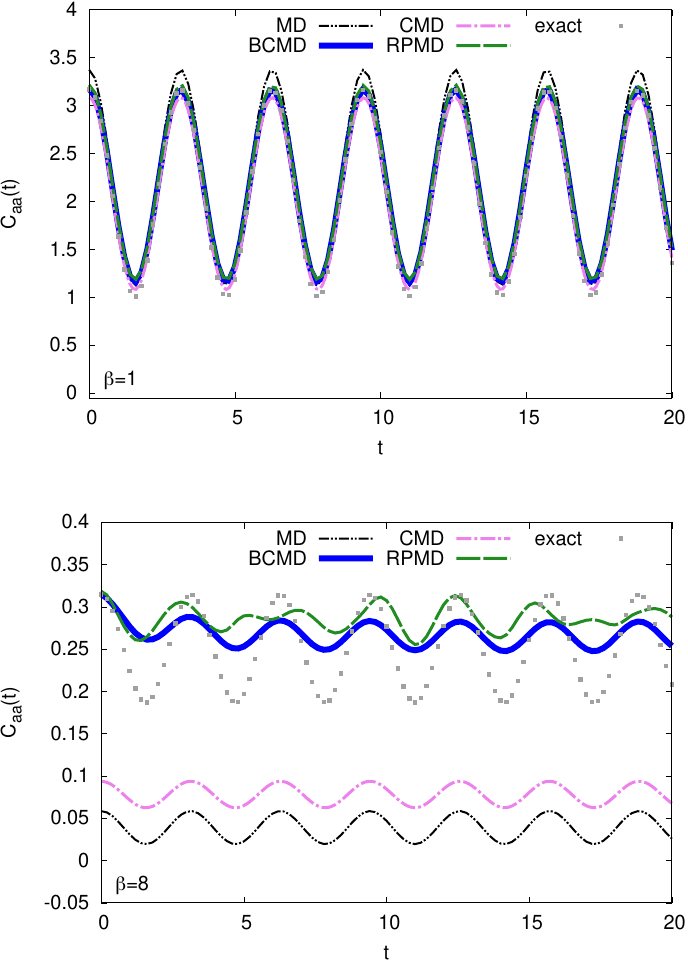}
\end{center}
\caption{Shiga, submitted to JCC.}
\label{fig3}
\end{figure}

\begin{figure}[htbp]
\begin{center}
\includegraphics[width=1.0\linewidth]{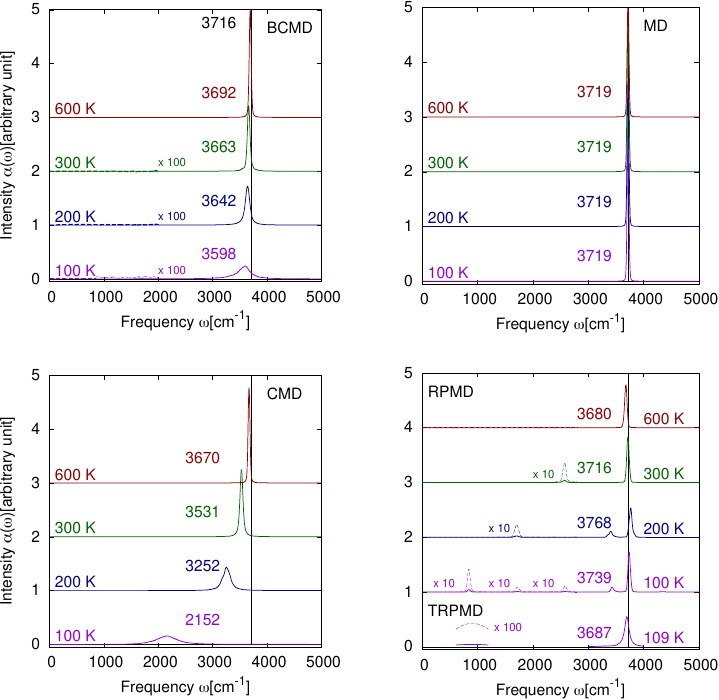}
\end{center}
\caption{Shiga, submitted to JCC.}
\label{fig4}
\end{figure}

\begin{figure}[htbp]
\begin{center}
\includegraphics[width=1.0\linewidth]{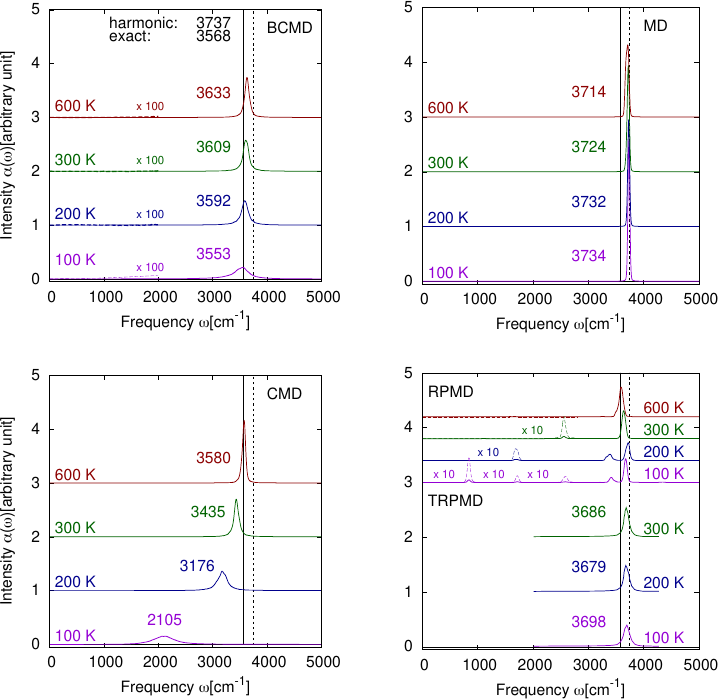}
\end{center}
\caption{Shiga, submitted to JCC.}
\label{fig5}
\end{figure}

\begin{figure}[htbp]
\begin{center}
\includegraphics[width=1.0\linewidth]{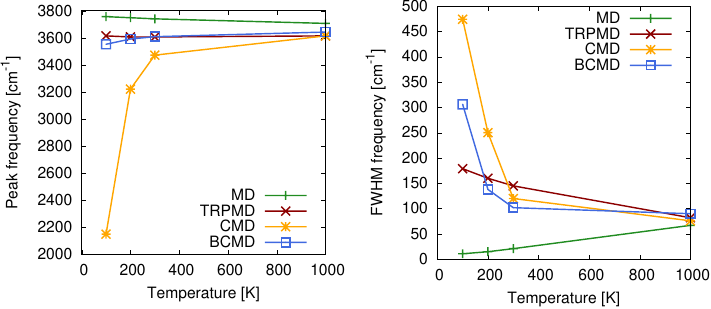}
\end{center}
\caption{Shiga, submitted to JCC.}
\label{fig6}
\end{figure}

\begin{figure}[htbp]
\begin{center}
\includegraphics[width=1.0\linewidth]{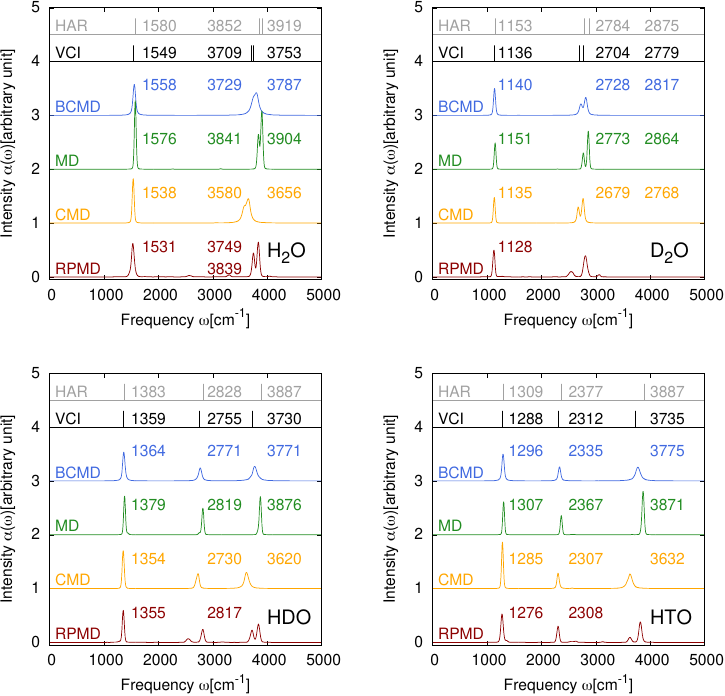}
\end{center}
\caption{Shiga, submitted to JCC.}
\label{fig7}
\end{figure}

\begin{figure}[htbp]
\begin{center}
\includegraphics[width=1.0\linewidth]{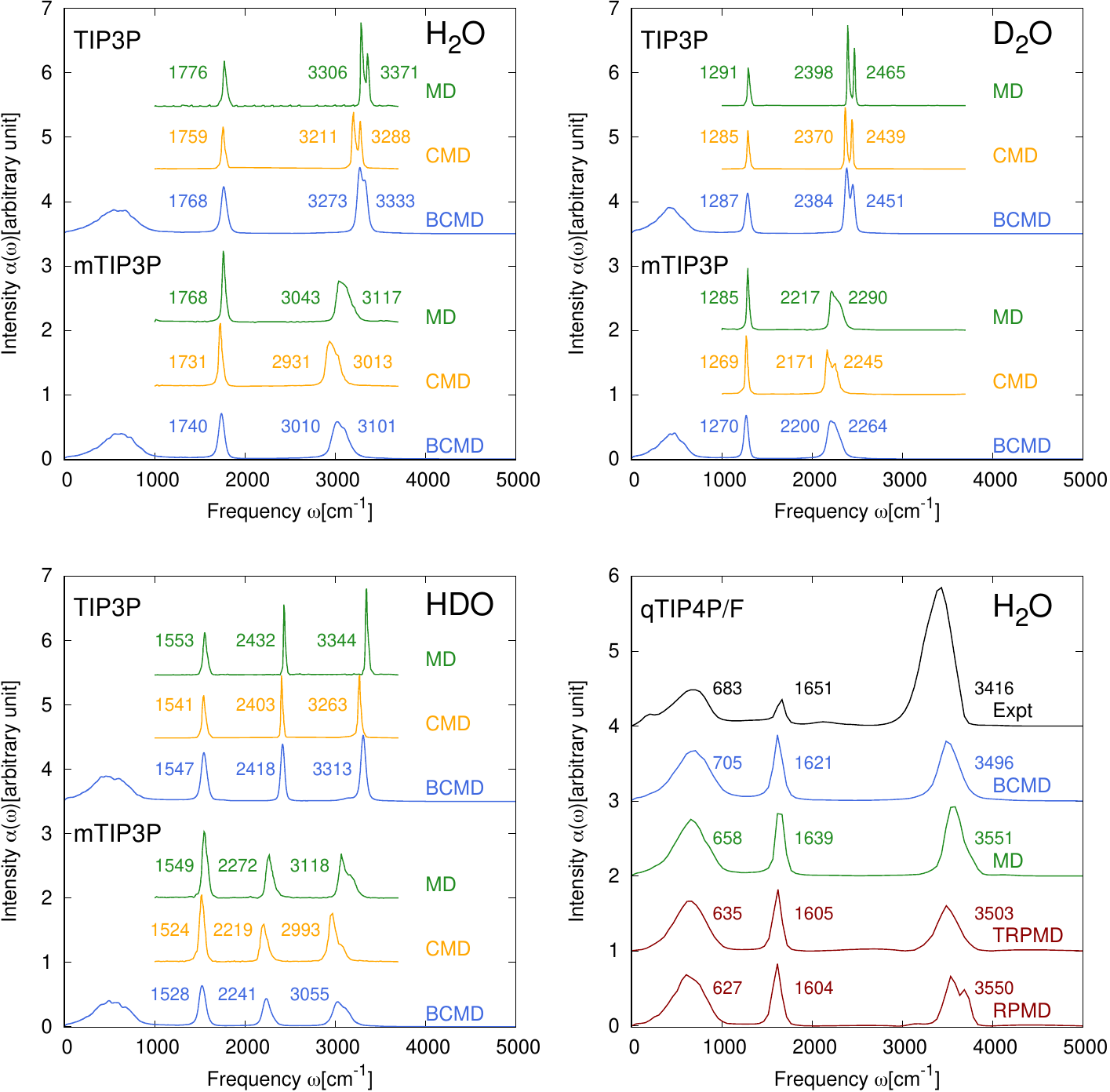}
\end{center}
\caption{Shiga, submitted to JCC.}
\label{fig8}
\end{figure}

\begin{figure}[htbp]
\begin{center}
\includegraphics[width=0.60\linewidth]{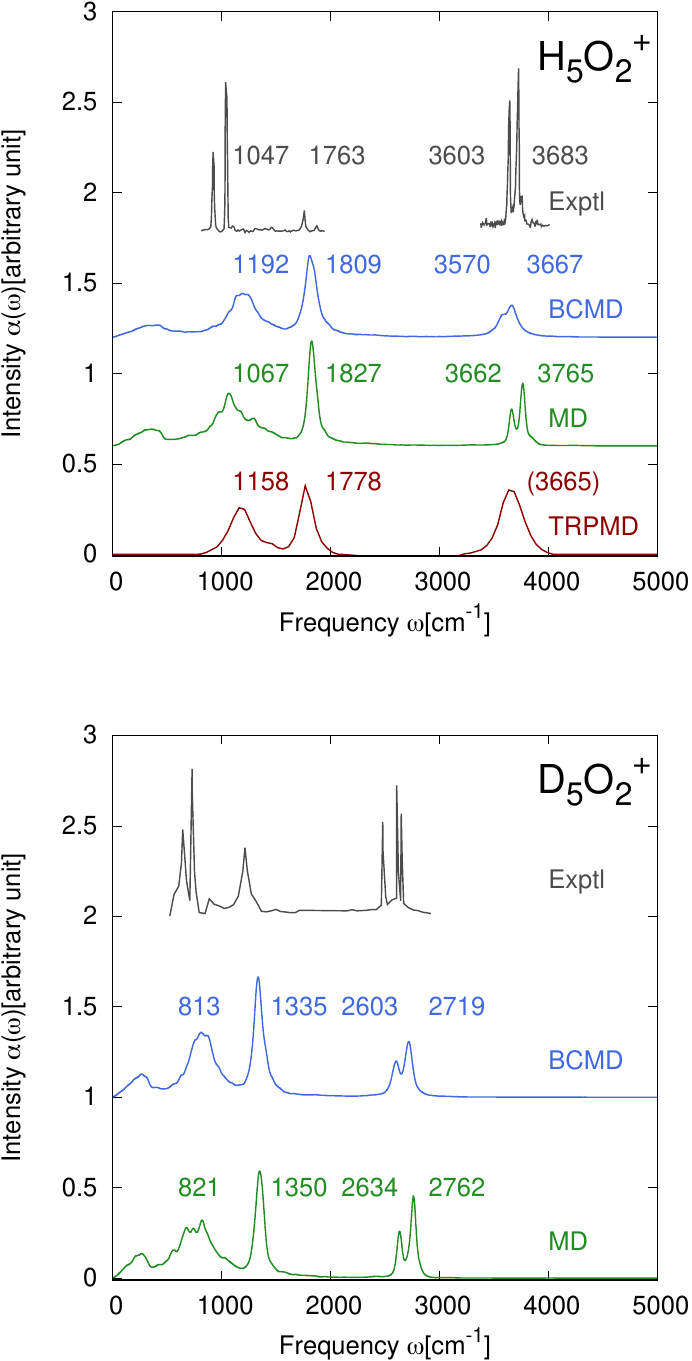}
\end{center}
\caption{Shiga, submitted to JCC.}
\label{fig9}
\end{figure}

\begin{figure}[H]
\begin{center}
\includegraphics[width=1.0\linewidth]{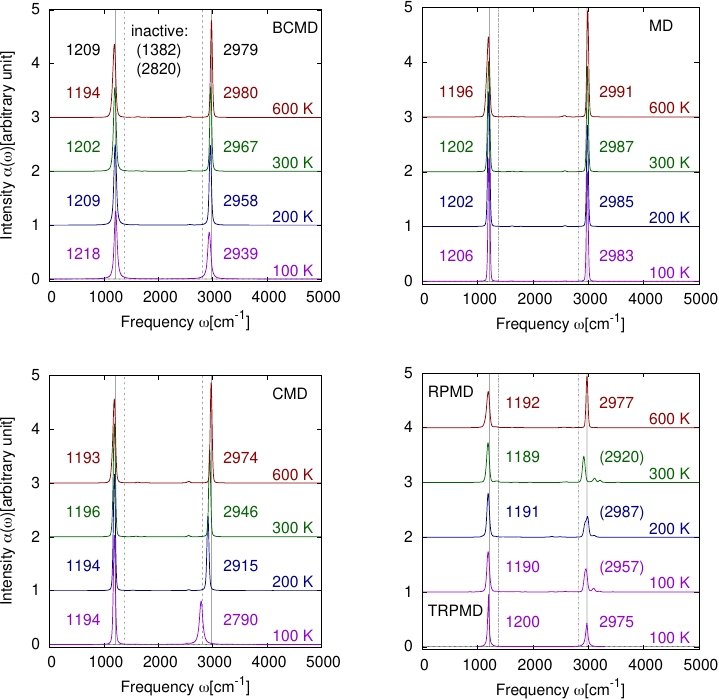}
\end{center}
\caption{Shiga, submitted to JCC.}
\label{fig13}
\end{figure}

\begin{figure}[H]
\begin{center}
\includegraphics[width=1.0\linewidth]{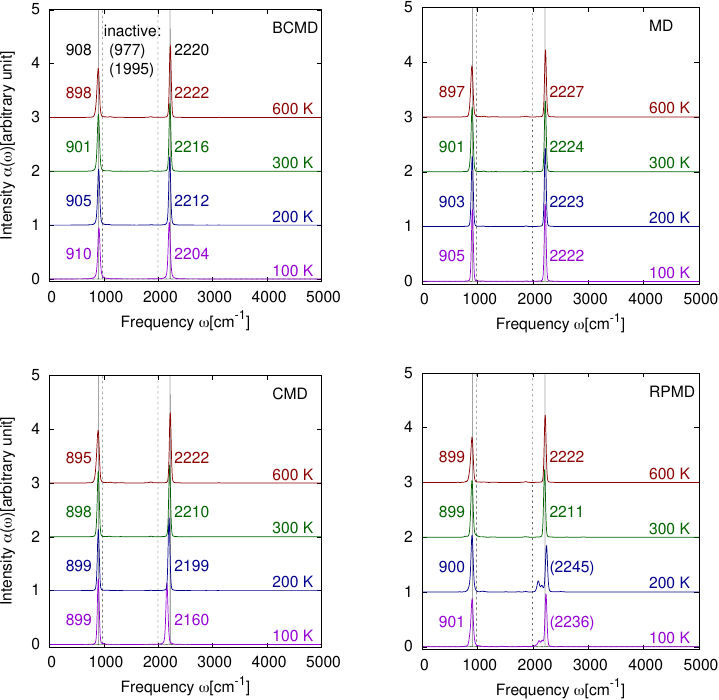}
\end{center}
\caption{Shiga, submitted to JCC.}
\label{fig14}
\end{figure}

\begin{itemize}
\item{Figure \ref{fig1}:
 Kubo-transformed correlation function $C_{xx}^{}(t)$ of
 a mild anharmonic potential,
 $V(x) =\frac{1}{2} x^2 + \frac{1}{10} x^3 + \frac{1}{10} x^4$
 with $m=1$ and $\hbar=1$.
The results are shown for BCMD (solid blue lines),
 MD (black double-dotted-dashed lines),
 CMD (violet single-dotted-dashed lines),
 RPMD (green dashed lines), and the exact results
 $C_{xx}^{\rm kubo}(t)$ (grey dotted lines) at
 a high temperature, $\beta = 1$ (top panel)
 and at a low temperature, $\beta = 8$ (bottom panel).
 The number of beads is set to $P = \frac{1}{4\beta}$ for BCMD\@.
}
\item{Figure \ref{fig2}:
 Kubo-transformed correlation function $C_{xx}^{}(t)$ of
 a quartic potential, $V(x) =\frac{1}{2} x^4$,
 otherwise the same as Figure \ref{fig1}.
}
%
\item{Figure \ref{fig3}:
 Kubo-transformed correlation function $C_{aa}^{}(t)$
 with $a = x^2$ of a harmonic potential
 $V(x) = \frac{1}{2} x^2$,
 otherwise the same as Figure \ref{fig1}.
 For CMD, the result of the effective classical operator
 \cite{jang1999derivation,liu2011two} is shown.
}
%
\item{Figure \ref{fig4}:
 Vibrational spectra of OH molecule obtained from the BCMD (top left),
 MD (top right), CMD (bottom left), RPMD (bottom right), and
 optimally-damped TRPMD (bottom right) methods
 for harmonic model, $V=\frac{K}{2}(r-R)^2$, where
 $r$ is the OH bond length.
 The parameters are $K=0.49536$
 hartree$\cdot$bohr$^{-2}$ and $R = 1.8897$ bohr.
 The black line represents the exact frequency of
 $(v,r) = (0,0) \rightarrow (1,0)$ transition of this model.
 The dotted lines in BCMD, RPMD and TRPMD are the magnified views.
 }
%
\item{Figure \ref{fig5}:
 Vibrational spectra of OH molecule obtained in the same way
 as Figure \ref{fig4}, but for Morse model,
 $V = D \left(1-e^{-A(r-R)}\right)^2$,
 where $r$ is the OH bond length.
 The parameters are $R=1.8324$ bohr,
 $D=0.1875$ hartree, and $A=1.1562$ bohr$^{-1}$.
 The black solid line represents the exact frequency of
 $(v,r) = (0,0) \rightarrow (1,0)$ transition, while
 the black dashed line represents the harmonic frequency
 of this model.
 The results of TRPMD using the same Morse model were
 reproduced from Reference \cite{rossi2014remove}.
 The dotted lines in BCMD and RPMD are the magnified views.
}
\item{Figure \ref{fig6}:
 The temperature dependence on the position (left) and the FWHM
 value (right) of the spectral peak shown in Figure \ref{fig5}.
 The results of TRPMD using the same Morse model were
 reproduced from Reference \cite{rossi2014remove}.
}
%
\item{Figure \ref{fig7}:
 Vibrational spectra of gaseous H$_2^{}$O (top-left),
 D$_2^{}$O (top-right), HDO (bottom-left),
 and HTO (bottom-right) at 300 K using the q-TIP4P/F model.
 The results of BCMD, MD, CMD, RPMD are shown along with
 the frequencies of $(v,r) = (0,0) \rightarrow (1,0)$
 transitions obtained by the VCI calculations,
 and those in the harmonic approximation (HAR).
 The peak positions were obtained by fitting the spectrum to
 the sum of Lorentzian functions.
}
%
\item{Figure \ref{fig8}:
 Vibrational spectra of liquid H$_2^{}$O (top-left),
 D$_2^{}$O (top-right) and HDO (bottom-left) obtained from the MD
 and BCMD methods
 using the TIP3P and mTIP3P models, and
 vibrational spectra of liquid H$_2^{}$O
 obtained from the MD and BCMD methods
 using the q-TIP4P/F model (bottom-right).
 The results of the CMD method
 were reproduced from Reference \cite{ivanov2010communications}, and
 the results of the RPMD and the optimally-damped TRPMD methods
 were reproduced from Reference \cite{rossi2014remove}.
 The experimental infrared spectrum was reproduced from
 Reference \cite{bertie1996infrared}.
 The peak positions were obtained by fitting the spectrum to
 the sum of Lorentzian functions.
}
%
\item{Figure \ref{fig9}:
 The infrared spectra of H$_5^{}$O$_2^+$ (top) and
 D$_5^{}$O$_2^+$ (bottom) obtained from the ab initio MD
 and BCMD methods at 300 K\@.
 The peak positions were obtained by fitting the spectrum to
 the sum of Lorentzian functions.
 The result of the optimally-damped TRPMD methods
 using ab initio-based force field
 were reproduced from Reference \cite{rossi2014remove}.
 The experimental results were reproduced from
 Reference \cite{guasco2011unraveling},
 for those of H$_5^{}$O$_2^+\cdot$Ne and D$_5^{}$O$_2^+\cdot$Ar.
}
%
%
%
%
\item{Figure \ref{fig13}:
 Vibrational spectra of CH$_4^{}$ molecule obtained from the
 BCMD (top left), MD (top right), CMD (bottom left),
 RPMD and optimally-damped TRPMD (bottom right) methods
 using the harmonic force field
 in Reference \cite{witt2009applicability}.
 The black lines represent the harmonic frequencies of
 this model for the infrared active modes (solid) and
 inactive modes (dotted).
}
%
\item{Figure \ref{fig14}:
 The same as Figure \ref{fig13}, but for CD$_4^{}$ molecule.
}
\end{itemize}

\clearpage
%
%
\clearpage
%
%
\clearpage
%
%
\clearpage
%
%
\clearpage
%
%
\clearpage
%
%
\clearpage
%
%
\clearpage
%
%
\clearpage
%
%
\clearpage
%
%
\clearpage
%
%
\clearpage
%
%
\clearpage
%
%
\clearpage
%
%

\end{document}